\documentclass[final,5p,times,twocolumn]{elsarticle}

\usepackage{natbib}

\usepackage{amssymb}
\usepackage{lipsum}
\usepackage{listings}
\usepackage{amsmath}
\usepackage{hyperref}
\usepackage{cleveref}

\makeatletter 
\def\ps@pprintTitle{%
 \let\@oddhead\@empty
 \let\@evenhead\@empty
 \let\@evenfoot\@oddfoot} % Supprimer le bas de page ELSEVIER
\makeatother

\begin{document}

\begin{frontmatter}

\title{AdlerPy: A Python Package for the Perturbative Adler Function}

\author[]{Rodolfo Ferro-Hernández \small{\\email: \texttt{rferrohe@uni-mainz.de}}
\\\small{\textit{Institute for Nuclear Physics,
Johannes Gutenberg University, 55099 Mainz, Germany.}}}

\begin{abstract}
In this letter, I give availability to the source code AdlerPy which will allow to easily use the Adler function. As an application, I use the mass-dependent perturbative expression for the Adler function to compute several observables and relevant Standard Model parameters.     
\end{abstract}


\end{frontmatter}

\section{Introduction}
\label{introduction}
The Adler function, Ref.~\cite{Adler:1974gd}, is defined as the logarithmic derivative of the electromagnetic correlator $\Pi(Q^2)$, namely
\begin{equation}
D(Q^2) = -12\pi^2Q^2\frac{d\Pi(-Q^2)}{dQ^2},
\end{equation}
where 
\begin{equation}
\left(-q^2 \eta^{\mu\nu}+q^{\mu} q^{\nu} \right)\Pi(q^2) = i\int d^4x e^{iqx}\langle 0|TJ_{em}^{\mu}(x)J_{em}^{\nu}(0)|0\rangle,
\end{equation}
and $Q^2 = -q^2$. $J_{em}$ is the electromagnetic current including the quark charges. The Adler function is a relevant quantity used to analyze several observables. Among others, it is used to study the cross-section of $e^{+}e^{-}\rightarrow\mathrm{hadrons}$, the anomalous magnetic moment of the muon, the running of the electromagnetic coupling constant, the hadronic $\tau$ and $Z$ decays, or to compare with predictions from direct calculations using a discretized Euclidean spacetime (lattice QCD). It can also be used to complement the results of the future MUonE, Ref.~\cite{Abbiendi:2022oks}, experiment. Given its dependence on the quark masses, at a given order in the strong coupling constant $\hat{\alpha}_s$, the massive contribution to the Adler function is known partially, \textit{i.e.}, in specific kinematical regions. At orders $\hat{\alpha}^{0}_s$ and $\hat{\alpha}^{1}_s$, it is known exactly for all $Q^2$. At order $\hat{\alpha}^2_s$, the expansion for low $Q^2$ is written as a Taylor series in terms of the parameter ($Q^2/m^2$), where $m$ is the mass of the corresponding quark. The expansion is known with up to 30 terms, Ref.~\cite{Maier:2007yn,Maier:2011jd}. The same number of terms are known at this order in the high-energy regime, where the expansion parameter is ($m^2/Q^2$), Ref.~\cite{Maier:2011jd}. There are also terms coming from the so-called double bubble diagrams, where the mass of the inner and outer quark bubbles are different. The expressions for the contribution of these diagrams to the cross-section of $e^+e^-\rightarrow\mathrm{hadrons}$ (the so-called $R$-ratio after normalization) are given in Refs.~\cite{Kniehl:1989kz,Hoang:1994it,Harlander:2002ur}. The results for $R$ can be converted to the Adler function using an integral expression
\begin{equation}
D(s) = Q^2\int^{\infty}_{0}\frac{R(s)}{\left(s+Q^2\right)}ds. \label{eq:AdlerfromR}
\end{equation}
as shown in Ref.~\cite{Davier:2023hhn}. 

At order $\hat{\alpha}^3_s$, the knowledge is more limited. In the low-energy expansion, only terms up to $(Q^2/m^2)^{3}$ are known, while higher orders can be estimated, a common way to do it is with Padé approximants, for example. In the high-energy limit, the term up to order $(m^2/Q^2)$ was obtained in Ref.~\cite{Baikov:2004ku}. Reconstructions of the vacuum polarization function (and hence the Adler function) are given in Ref.~\cite{Maier:2017ypu}. At order $\hat{\alpha}^4_s$,  mass-less results are available, namely the first term in the expansion $(m^2/Q^2)$.

There are some codes in the literature that allow to compute the perturbative contribution to the Adler function. For example, one may use the code \texttt{rhad} written in Fortran, Ref.~\cite{Harlander:2002ur}, which computes the ratio to then perform the integral in \cref{eq:AdlerfromR}
But this involves going through an integral as an intermediate step. Another code available in the literature, which does not go through the integral, is \texttt{pQCDAdler} written in Fortran and provided by Ref.~\cite{Jegerlehner:2008rs},  AdlerPy's primary aim is to provide an easily accessible library for computing perturbative contributions to the Adler function, with Python chosen as the source code language for improved accessibility and usability, sacrificing (as usual in Python) performance. In the following sections I will explain how to use the code. As we will see, the perturbative corrections when the pole masses are used are big, as I will show in the examples of this letter.

\section{Installation and Code Snippets}

The code can be either downloaded from the GitHub repository  \url{https://github.com/rodofer2020/adlerpy}. or directly through the pip install command in Python, by simply typing the following in the terminal:
\begin{center} 
\texttt{pip install adlerpy}
\end{center} 
For the running of $\hat{\alpha}_s$, the code relies on \texttt{rundec}, while for interpolation and special functions, the standard Python libraries \texttt{numpy}, \texttt{mpmath}, and \texttt{scipy} are used. If the user does not have them installed, the previous pip install command will automatically install them. 

The source code consists of two main files. One is called \texttt{adler\_routines}, which contains the expansions provided in the references mentioned in the introduction, without referring to a specific quark. If the reader is interested in understanding exactly what is contained in a specific routine, the Python command \texttt{help} will provide such information. I have added an example of such information in the Appendix.

The other important file is called \texttt{adler\_sm}, which combines the functions defined in \texttt{adler\_routines} to provide the Standard Model prediction. For example, the routine \texttt{adler\_charm\_pert} computes the charm quark contribution to the Adler function. It is in this routine where the interpolation between low and high energy is done, or where the number of active quarks is properly included. In the following sections, I will show different applications of the code. The examples are also provided as Jupyter notebooks in the same GitHub repository.

\section{Comparison between Pole and $\overline{\mathrm{MS}}$ Masses}
One interesting application is the study of the differences between the use of the pole and the $\overline{\mathrm{MS}}$ definitions of the heavy quark masses in the Adler function. It is known that the pole mass definition contains renormalons, which makes the perturbative relation between both ill-behaved. Furthermore, it has been pointed out in recent years that the pole mass definition should be abandoned due to this problematic behavior. On the other hand, if the observable in question involves heavy quarks that are close to being on-shell, the $\overline{\mathrm{MS}}$ may also lead to large perturbative corrections. This suggests that new definitions of the mass must be taken which are numerically close to the on-shell mass but that are renormalon-free. See, for example, Ref.~\cite{Beneke:2021lkq}. Since for the Adler function, the quarks are not on-shell, like in direct production processes, the $\overline{\mathrm{MS}}$ definition is expected to yield a stable result. To confirm such a statement, a comparison with the lattice prediction is done, which is the application of this section.

\begin{figure}[h]
    \centering
    \includegraphics[width=1\linewidth]{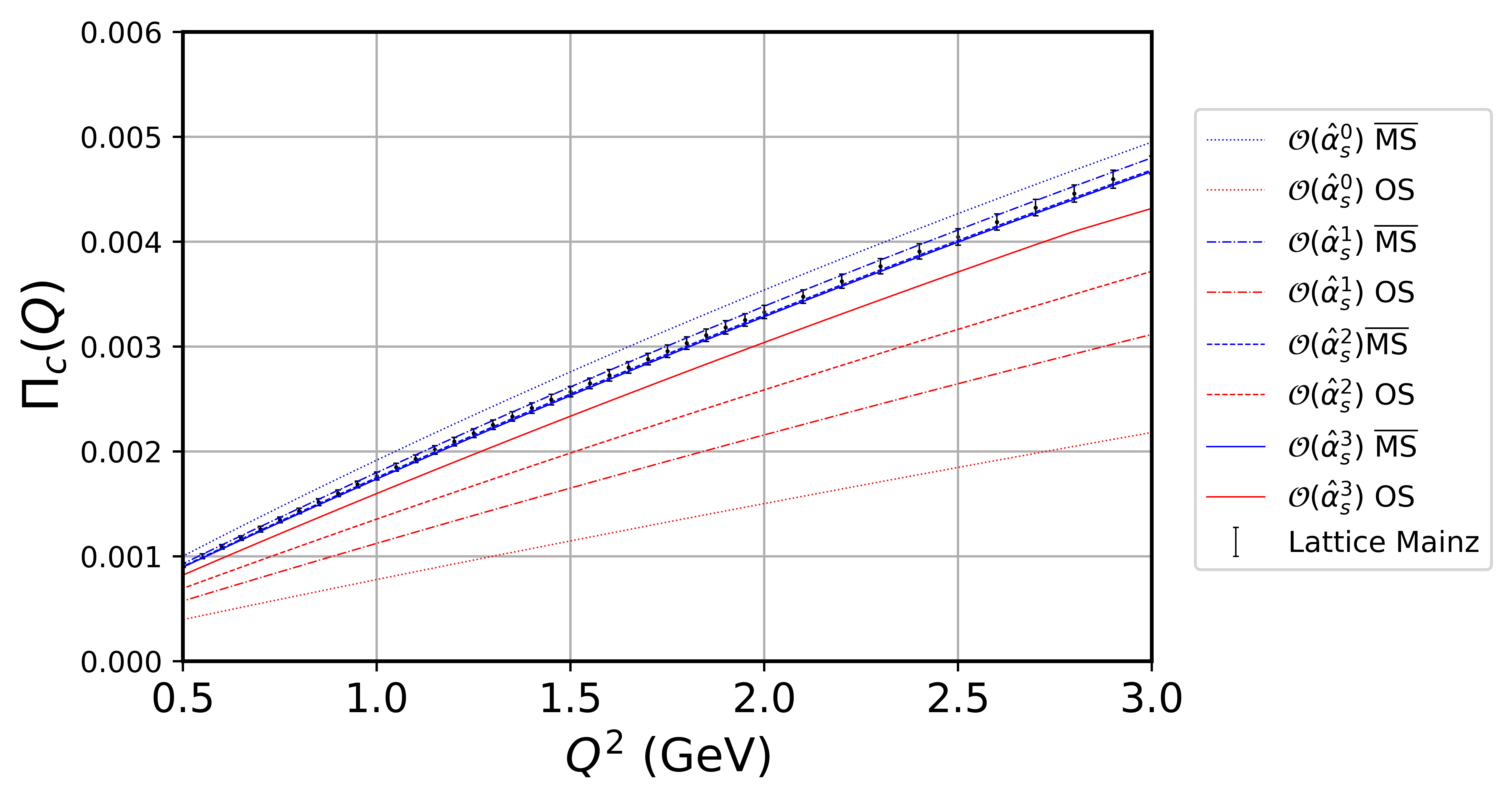}
    \caption{Comparison of the subtracted charm vacuum polarization function between lattice QCD (black points) and the $\overline{\mathrm{MS}}$ (blue) and on-shell schemes (red) at different orders in perturbation theory, as computed using Eq.~\cref{eq:subtractedfromD}.}
    \label{fig:substractedlatcharm}
\end{figure}

The subtracted vacuum polarization function can be computed using
\begin{equation}
    \Pi(0) - \Pi(-Q^2) = \frac{1}{12\pi^2}\int^{Q^2}_{0}\frac{D(Q^2)}{Q^2}\,dQ^2\label{eq:subtractedfromD}.
\end{equation}
Ignoring disconnected diagrams, where the quark flavor may change, the vacuum polarization can be split into different flavor contributions. 
In particular, in this section, I will be interested in the connected contributions coming from the charm quark. Such a term is contained in the routine \texttt{adler\_charm\_pert}. This routine contains the parameter \texttt{mpole\_on}, which, if set to \texttt{true}, will use the pole mass expansion for the Adler function, while the order in perturbation theory can be controlled through the parameter \texttt{nloops}.  For this comparison, I will use the PDG values for the pole and $\overline{\mathrm{MS}}$ masses, namely $M_c = 1.67\,\mathrm{GeV}$ and $\hat{m}_c(\hat{m}_c) = 1.273\,\mathrm{GeV}$, as well as for the strong coupling constant $\hat{\alpha}_s(M_Z) = 0.1185$. 
As the control measurement, one can use the lattice results from the Mainz collaboration, Ref.~\cite{Ce:2022eix}, which give the vacuum polarization function of the charm quark for several values of $Q^2$. The results can be seen in Fig.~\ref{fig:substractedlatcharm}.

From Fig.~\ref{fig:substractedlatcharm}. one can see that the $\overline{\mathrm{MS}}$ pQCD prediction is more stable when higher orders are included. This, as mentioned before, seems to be a consequence of the renormalons which are included in the definition of the pole mass. Furthermore, it is in good agreement with the lattice prediction.  From this observation, we can conclude that is more convenient to use the $\overline{\mathrm{MS}}$ mass to reduce the perturbative uncertainty. 

\begin{figure}[h]
    \centering
    \includegraphics[width=1\linewidth]{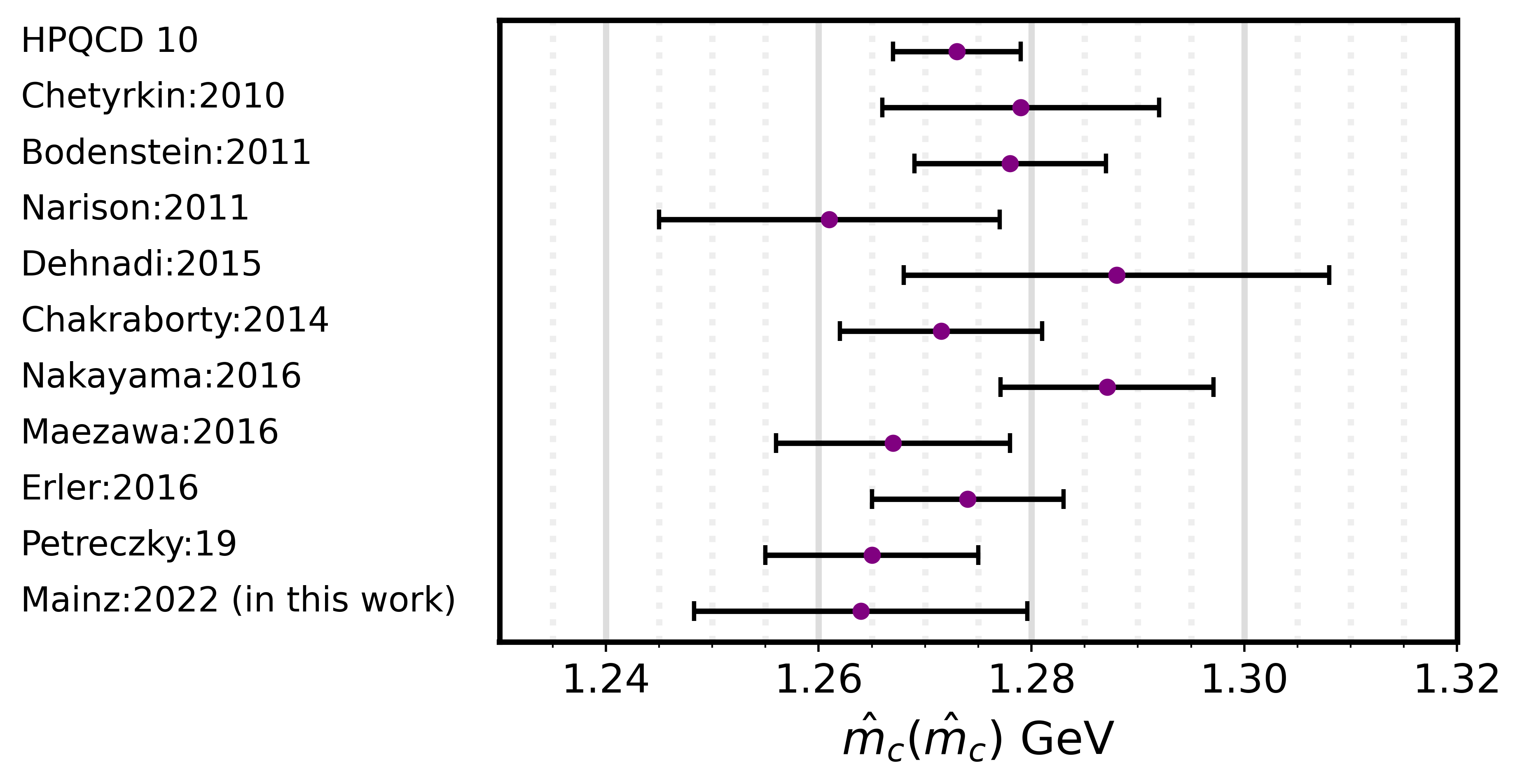}
    \caption{Comparison of the $\overline{\mathrm{MS}}$ charm mass extracted from the lattice QCD results of \cite{Ce:2022eix} with other determinations of the mass \cite{Chetyrkin:2010ic, Bodenstein:2011ma, Narison:2011xe, Dehnadi:2015fra, Chakraborty:2014aca, Nakayama:2016atf, Maezawa:2016vgv, Erler:2016atg, Petreczky:2019ozv}. }
    \label{fig:charmmasscomparison}
\end{figure}

As a further result, it is interesting to see to what value of $\hat{m}_c\left(\hat{m}_c\right)$ the lattice result corresponds to. Unfortunately, Ref.~\cite{Ce:2022eix} does not give a point-by-point correlation for the charm vacuum polarization function. Nevertheless, they do give the correlation in a parametric way for the total vacuum polarization, which shows a large correlation for points that are nearby, as shown in Ref.~\cite{Davier:2023hhn}. This feature can also be seen in Fig.~\ref{fig:substractedlatcharm}, since there seems to be no "randomness" in the lattice points. Hence, one may assume a 100\% correlation between points and take just one of them to extract the mass. We obtain that the corresponding mass is 
\begin{align}
    \hat{m}_c\left(\hat{m}_c\right) &= 1.263 + 5.6\left(\hat{\alpha}_s(M_Z) - 0.1185\right)\nonumber \\
    &\pm 0.012_{\mathrm{lat}}\pm 0.005_{\mathrm{tr}}\pm 0.002_{\mathrm{cond}}\,\mathrm{GeV}.
\end{align}
Which is in agreement with other determinations of the charm quark mass. See Fig.~\ref{fig:charmmasscomparison}.

\section{The heavy quark contribution to the anomalous magnetic moment of the muon}
%%\label{}
The anomalous magnetic moment of the muon, denoted as $a_\mu$, has been a highly relevant observable in recent decades. There is a well-known tension between the Standard Model prediction, as calculated using $e^{+}e^{-}\rightarrow \mathrm{hadrons}$ data \cite{Aoyama:2020ynm}, and the measured value in Ref.~\cite{Muong-2:2023cdq}. This tension has prompted new physics explanations for many years. However, a recent publication from the BMW lattice QCD collaboration, has provided a prediction with a competitive error \cite{Borsanyi:2020mff}. Their result align with the experimental value, raising new questions about the original discrepancy.

The value of $a_\mu$ can be computed from the Adler function using the relation:

\begin{equation}
a_{\mu}=\left(\frac{\alpha}{\pi}\right)^2 \int^{\infty}_{0}\frac{\left(1-u(Q^2)\right)^2}{\left(1+u(Q^2)\right)^2}\frac{D\left(Q^2\right)}{6 Q^2}dQ^2,
\end{equation}where $u(Q^2)=\sqrt{1+\frac{4m^2_\mu}{Q^2}}$, which can  be derived from the expression given by Ref. \cite{Blum:2002ii}. For additional relations involving the Adler function, readers are encouraged to refer to Ref. \cite{Nesterenko:2021byp}.

As an illustrative example, I computed the contribution to the anomalous magnetic moment of the muon from charm (denoted as $a^{(c)}_\mu$) and bottom (denoted as $a^{(b)}_\mu$) quarks. While these contributions do not dominate the total value or the error, it is prudent to minimize uncertainties in the Standard Model prediction. The charm contribution was computed using the routine \texttt{adler\_charm\_pert}, with input parameters $\hat{\alpha}_s(M_Z)=0.1185\pm0.0016$, $\hat{m}_c\left(\hat{m}_c\right)=\left(1.273\pm0.008\right)\mathrm{GeV}$, and $\hat{m}_b\left(\hat{m}_b\right)=\left(4.18\pm0.008\right)\mathrm{GeV}$. It should be noted that there exists a correlation between the value of $\hat{\alpha}_s$ and the central value of $\hat{m}_c\left(\hat{m}_c\right)$, as discussed in Ref. \cite{Erler:2016atg}. The result for $a^{\left(c\right)}_\mu$ is:

\begin{equation}
a^{\left(c\right)}_\mu=\left(14.57\pm0.12_{\alpha_s}\pm0.20_{m_c}\pm0.1_{\mathrm{tr}}\right)\times10^{-10}.
\end{equation}

Here, the truncation error is defined as the difference between including or not the $\hat{\alpha}^3_s$ moments. In the code, this is achieved by changing the parameter \texttt{nloops} from 3 to 2. The $\alpha^2$ QED effects amount to $0.02\times10^{-10}$ and are already included in our result. The inclusion or omission of these terms is controlled by setting the parameter \texttt{QED} to either \texttt{True} or \texttt{False}, respectively. Notably, this result aligns well with the sum rules approach quoted in Ref. \cite{Erler:2000nx, Kennedy:2021ysp}, in particular is remarkable the agreement with Mainz lattice calculation Ref. \cite{Gerardin:2019rua},

\begin{equation}
a^{\left(c\right)}_\mu=\left(14.66\pm0.45_{\mathrm{stat}}\pm0.06_{\mathrm{sys}}\right)\times10^{-10},
\end{equation}

or the results of the BMW collaboration (Ref. \cite{Budapest-Marseille-Wuppertal:2017okr, Borsanyi:2020mff}):

\begin{equation}
a^{\left(c\right)}_\mu=\left(14.6\pm0.0_{\mathrm{stat}}\pm0.1_{\mathrm{sys}}\right)\times10^{-10}.
\end{equation}

Applying a similar approach to the bottom quark, we obtain:

\begin{equation}
a^{\left(b\right)}_\mu=\left(0.302\pm0.001_{\alpha_s}\pm0.001_{m_c}\pm0.001_{\mathrm{tr}}\right)\times10^{-10}.
\end{equation}

This result also aligns well with the results obtained using sum rules by Ref. \cite{Erler:2000nx, Kennedy:2021ysp}.

\section{Extraction of $\hat{\alpha}_s$ from lattice data}
The strong coupling constant $\hat{\alpha}_s$ is a key parameter in the Standard Model. A few years ago, it was proposed that it may be possible to extract it through a comparison between the Adler function obtained from lattice QCD and the one from pQCD \cite{Francis:2014yga, Hudspith:2016yhn}.

Recently, in Ref. \cite{Davier:2023hhn}, this analysis was done using the Mainz lattice results \cite{Ce:2022eix}. It was found that pQCD and lattice Adler functions have very good agreement with each other, while the one obtained from $e^+e^-\rightarrow\mathrm{hadrons}$ is in tension with both of them. This discrepancy is in turn connected to the well-known tension between lattice and experimental data in the calculation of the anomalous magnetic moment of the muon.

\begin{figure}[h]
    \centering
    \includegraphics[width=1\linewidth]{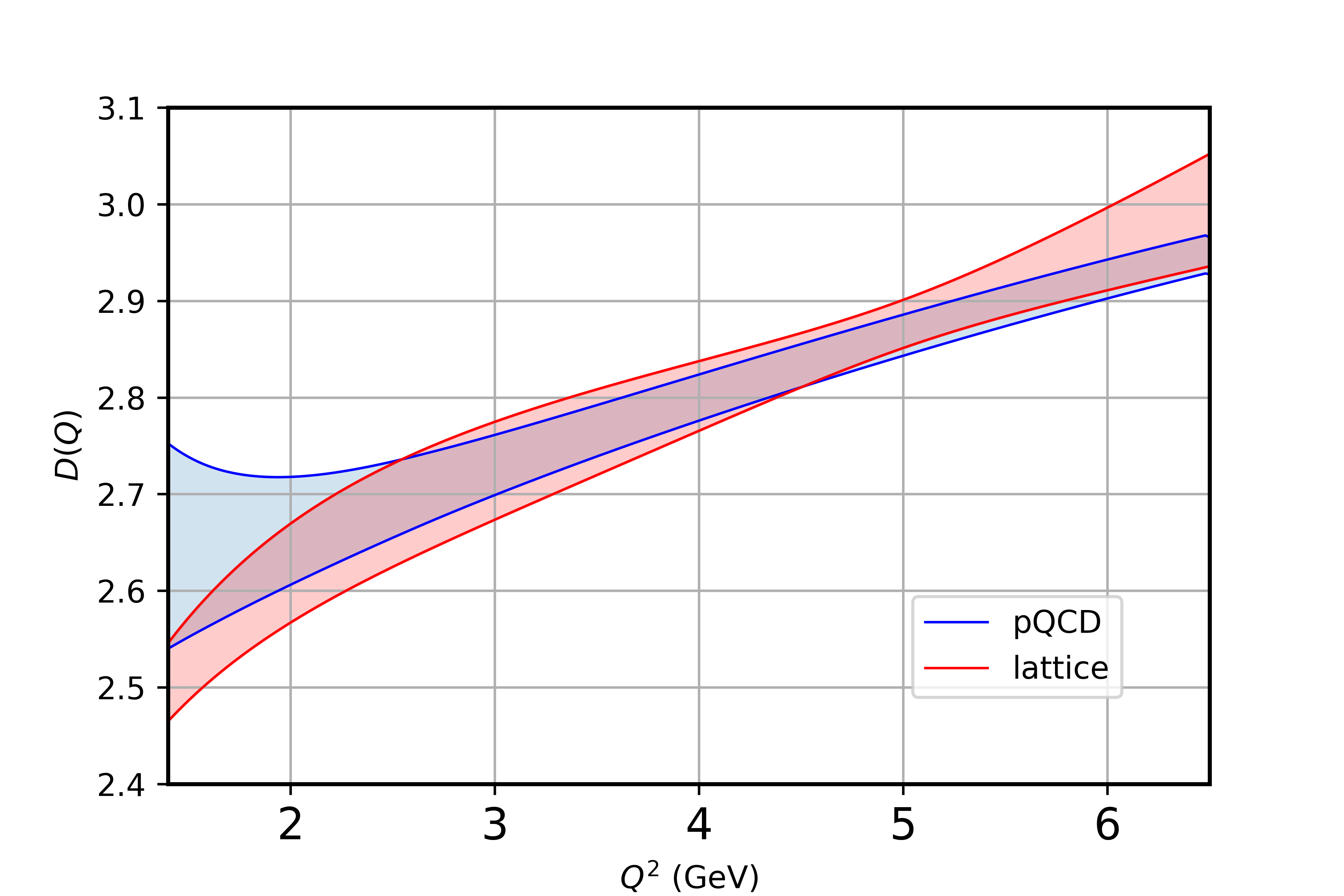}
    \caption{Comparison between the Adler function obtained from the Mainz lattice collaboration and pQCD. This plot is similar to the one obtained in Ref. \cite{Davier:2023hhn}. Here the parametric error is larger, taken to be $\hat{\alpha}_s = 0.1185 \pm 0.0016$.}
    \label{fig:AdlerlatticevspQCD1}
\end{figure}

As another example of the applications of the perturbative Adler function, we can redo the analysis shown in Ref. \cite{Davier:2023hhn} using the routines \texttt{adler\_light\_pert} and \texttt{adler\_charm\_pert}. The values of the condensates can be controlled through the variables \texttt{GG} and \texttt{qq}. Only the dimension-four contributions are included, for which I use the expressions given in Ref. \cite{Davier:2023hhn}. The results can be seen in Fig. \cref{fig:AdlerlatticevspQCD1}. As expected, we recover the good agreement between lattice and pQCD. From here, one can easily extract the strong coupling constant. Since the prediction for the Adler function from lattice QCD has a big point-by-point correlation, I compute $\hat{\alpha}_s$ from the Adler using just one point, i.e., $Q=2\,\mathrm{GeV}$, and obtain\footnote{This result is slightly larger than the one shown in Ref.~\cite{Davier:2023hhn}. The reason for this is that I do not include the bottom quark since the lattice result does not include them. Furthermore, in Ref. ~\cite{Davier:2023hhn}, there is an estimation for the higher-order terms.}

\begin{equation}
\hat{\alpha}_s(M_Z) = 0.1186 \pm 0.0030\quad\mathrm{From\,\,\, Mainz\,\,\, LQCD}.
\end{equation}

\begin{figure}[h]
    \centering
    \includegraphics[width=1\linewidth]{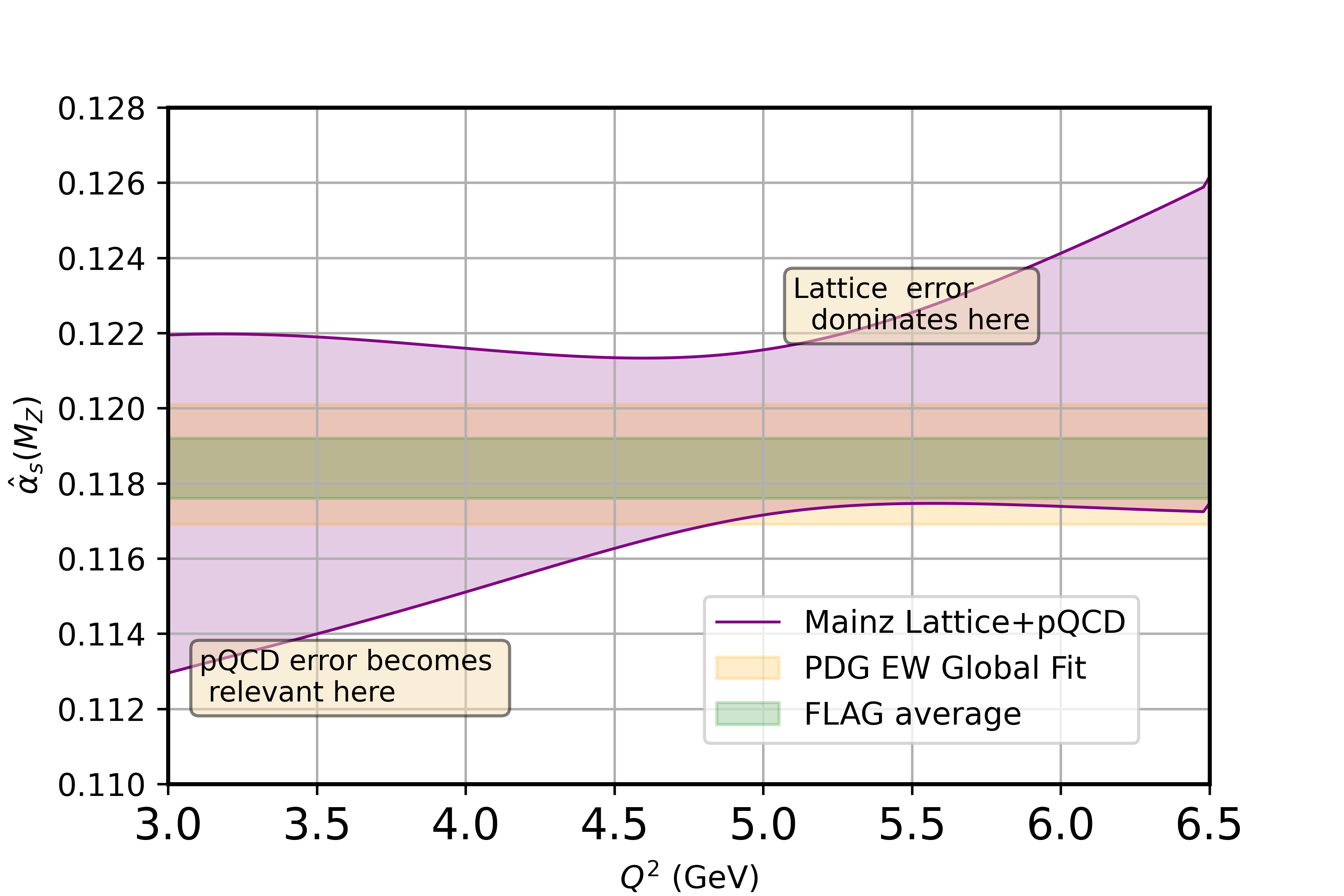}
    \caption{Extraction of $\hat{\alpha}_s(M_Z^2)$ from the Mainz lattice data. This plot is similar to the one obtained in Ref. \cite{Davier:2023hhn}.}
    \label{fig:AdlerlatticevspQCD2}
\end{figure}

A comparison of the extracted value from different values of $Q^2$ is given in Fig. \ref{fig:AdlerlatticevspQCD2}.

\section{Calculation of $\Delta\alpha(M^2_Z)$ from lattice}

Another important application is the calculation of the hadronic contribution to the running of the electromagnetic coupling constant at the Z scale. The hadronic contribution to the running is given by

\begin{equation}
\Delta\alpha^{(5)}(M^2_Z)=4\pi\alpha\mathrm{Re}\left(\Pi(0)-\Pi(M_Z^2)\right).
\end{equation}

With the increased precision in lattice QCD results [Ref. \cite{Ce:2022eix}, Ref. \cite{Budapest-Marseille-Wuppertal:2017okr}], it is now possible to use them as a low-energy input for the vacuum polarization function. Then, this result can be run up to the electroweak scale where it is a key parameter in the relation between observables. Recently, in Ref. \cite{Erler:2023hyi}, a comparison between different methods to compute such running was done, with particular emphasis on the method of renormalization group equations and matching conditions. Here we explain with a bit more detail the so-called Euclidean split technique, which was first introduced in Ref. \cite{Jegerlehner:2008rs} and used subsequently in Ref. \cite{Proceedings:2019vxr} and Ref. \cite{Ce:2022eix}. It consists on splitting the contributions to $\Delta\alpha^{(5)}(M^2_Z)$ as

\begin{align}
\Delta\alpha^{(5)}(M^2_Z)&=\Delta\alpha_{\mathrm{lat}}^{(5)}(-Q^2_0)+\frac{1}{12\pi^2}\int^{M_Z^2}_{Q^2_0}\frac{D(Q^2)}{Q^2}dQ^2\nonumber \\
&+\left[\Delta\alpha^{(5)}(M^2_Z)-\Delta\alpha^{(5)}(-M^2_Z)\right]_{\mathrm{pQCD}},\label{eq:EST}
\end{align}where $Q_0$ is a low-energy scale. It is defined as the lowest energy scale where there is agreement between data (either lattice or dispersive) and pQCD. It is clear from the previous section that a scale of around $Q_0\approx 2\,$ GeV is a good scale, because it is a sweet spot between lattice QCD results and pQCD. Therefore, the first term on the right-hand side of eq. (\ref{eq:EST}) comes from a non-perturbative technique such as lattice QCD, while the second and third can be computed entirely in perturbative QCD. Even more, as we have shown in the previous sections, there is a remarkable agreement between pQCD for the heavy quark contributions for all values of $Q_0$. This motivates me to compute the charm and bottom quark contributions entirely in pQCD.

This calculation can be done using the routines \texttt{adler\_charm\_pert}, \texttt{adler\_bottom\_pert}, \texttt{adler\_light\_pert}, and \texttt{adler\_OZI\_pert}. I obtain

\begin{align}
\Delta\alpha^{(c)}(M^2_Z)&=\Bigg[79.55-\frac{1.3}{\mathrm{GeV}^4}\langle\hat{a}_s G^2\rangle+221\,\left(\hat{\alpha}_s-0.1185\right)
\nonumber \\
&-20\,\left(\hat{m}_c\left(\hat{m}_c\right)-1.273\,\mathrm{GeV}\right)\pm0.13_{\mathrm{tr}}\Bigg]\times 10^{-4},
\end{align}which is in agreement with the results obtained in Ref. \cite{Erler:2023hyi}. For the bottom quark, I obtain

\begin{align}
\Delta\alpha^{(b)}(M^2_Z)&=\Bigg[12.79+18\,\left(\hat{\alpha}_s-0.1185\right)
\nonumber \\
&-1.3\,\left(\hat{m}_b\left(\hat{m}_b\right)-4.18\,\mathrm{GeV}\right)\pm0.01_{\mathrm{tr}}\Bigg]\times 10^{-4}.
\end{align}While for the light quarks, I get\begin{align}
\Delta\alpha^{(3)}(M^2_Z)&=\Delta\alpha^{(3)}(-4\,\mathrm{GeV})+\Bigg[126.03+\frac{3.5}{\mathrm{GeV}^4}\langle\hat{a}_s G^2\rangle
\nonumber \\
&+\frac{13.8}{\mathrm{GeV}^4}\langle m_s \overline{q}q\rangle+130\,\left(\hat{\alpha}_s-0.1185\right)\pm0.05_{\mathrm{tr}}\Bigg]\times 10^{-4},
\end{align}which yields the total result (after adding a tiny contribution from the disconnected diagrams)

\begin{align}
\Delta\alpha^{(5)}(M^2_Z)&=\Delta\alpha^{(3)}(-4\,\mathrm{GeV})+\Bigg[218.34+\frac{2.2}{\mathrm{GeV}^4}\langle\hat{a}_s G^2\rangle
\nonumber \\
&+\frac{13.8}{\mathrm{GeV}^4}\langle m_s \overline{q}q\rangle+368\,\left(\hat{\alpha}_s-0.1185\right)\pm0.14_{\mathrm{tr}}\Bigg]\times 10^{-4}.
\end{align}Result in agreement with the one found in previous works like Ref. \cite{Erler:2023hyi}.

\section{Summary and conclusions}

The Adler function is a fundamental concept with numerous applications. In this paper, I have reviewed some of its practical uses. I utilized it to extract the charm mass from lattice QCD, comparing, in the process, the pole and $\overline{\mathrm{MS}}$ expressions. This comparison revealed that the expressions in terms of the pole mass exhibit a larger perturbative error and less agreement with the lattice data. On the contrary, the $\overline{\mathrm{MS}}$ expressions show a remarkable agreement even for Euclidean momenta below the QCD scale. This can be already inferred from the fact that the perturbative expressions at low energy only contain logarithms of the form $\ln(\frac{\mu}{\hat{m}_c})$, where $\hat{m}_c$ is above the strong interaction scale. Using this result, I employed the perturbative expression of the Adler function to compute the contributions of charm and bottom quarks to the anomalous magnetic moment of the muon, with a remarkable agreement with lattice and other theoretical determinations. Furthermore, I also extracted the strong coupling constant from lattice QCD by comparing it with the perturbative expansion of the Adler function. After that, I used the Euclidean split technique method to calculate the hadronic contribution to the electromagnetic constant at the electroweak scale.

I provided an explanation of the AdlerPy code used to perform these calculations and detailed instructions on its installation and usage. It is evident that there are areas where improvements can be made, such as the utilization of more sophisticated interpolation techniques for the Adler function. In this work, I computed it using splines through the SciPy predefined method. An efficient implementation of this could potentially accelerate calculations. Another potential upgrade involves the inclusion of moment estimates, which were not addressed in this study. Nevertheless, I am releasing this version with the hope that it will prove to be a valuable tool for computing relevant observables within the particle physics community.

\section{Acknowledgments}

This work was supported by the German-Mexican research collaboration grant SP 778/4-1 (DFG), 278017 (CONACYT) and CONACYT CB-2017-2018/A1-S-13051.
I would like to extend my gratitude to the Instituto de Física de la UNAM for their generous hospitality and support during my stay where this work was finished. I am also thankful to J. Erler, E. Peinado, F. Hagelstein, M. Gorshteyn, and H. Spiesberger for their valuable comments and fruitful discussions.

%%\label{}

%\section*{Acknowledgements}
%I would like to thank 

\appendix

\section{Some Routines}

The provided Python code consists of several functions and a class aimed at calculating contributions to the Adler function. This section summarizes the main functionalities of the code:

\subsection*{Functions}

\texttt{adler\_massless\_connected}: This function computes the connected and massless contribution to the Adler function. It utilizes the massless expression of $R$. Parameters include \texttt{Q}, \texttt{mu}, \texttt{asmu}, \texttt{nq}, and an optional \texttt{nloops}.

\texttt{adler\_massless\_disconnected}: Similar to the previous function, this one calculates the disconnected and massless Adler function contribution using the same principles and parameters.

\texttt{adler\_he\_o\_heavy\_i\_massless\_Q\_suppressed\_MS}: This function evaluates massive contributions to the high-energy expansion of the connected Adler function contribution in the $\overline{\mathrm{MS}}$ scheme. It considers an external massive quark and internal massless quarks. The input parameters include \texttt{m}, \texttt{Q}, \texttt{mu}, \texttt{asmu}, \texttt{nq}, \texttt{k}, and an optional \texttt{nloops}.

\texttt{adler\_le\_o\_heavy\_i\_massless\_m\_suppressed\_MS}: This function computes massive contributions to the low-energy expansion of the connected Adler function in the MSbar scheme. It involves an external massive quark and internal massless quarks. Input parameters comprise \texttt{m}, \texttt{Q}, \texttt{mu}, \texttt{asmu}, \texttt{nq}, \texttt{k}, and an optional \texttt{nloops}.

\texttt{adler\_o\_massless\_i\_heavy}: This function determines double bubble contributions with a massless external quark and a massive internal quark. It takes \texttt{m}, \texttt{Q}, and \texttt{cut} as parameters.

\texttt{adler\_hq\_full\_zero\_loop}: This function calculates the full Adler function expression at order $\alpha^{0}_s$ for a given scale \texttt{Q} and particle mass \texttt{m}.

\texttt{adler\_hq\_full\_one\_loop\_MS}: This function computes the full contribution of the Adler function at order $\alpha^{1}_s$ in the MSbar scheme. The input includes \texttt{Q}, \texttt{mhat}, and \texttt{mu}.

\texttt{adler\_hq\_full\_one\_loop\_OS}: Similar to the previous function, this calculates the Adler function contribution at order $\alpha^{1}_s$, but in the ON-shell scheme. The parameters include \texttt{Q} and \texttt{m}.

\texttt{alphas}: This function calculates the strong coupling constant at a specified scale \texttt{mu}. It relies on the \texttt{rundec} code. Input parameters are \texttt{aZ}, \texttt{Mz}, \texttt{mu}, \texttt{particles}, \texttt{nq}, and an optional \texttt{nloops}.

Another set of routines are: \texttt{adler\_charm\_pert}, \texttt{adler\_bottom\_pert}, \texttt{adler\_light\_pert}, and \texttt{adler\_OZI\_pert}. These routines give the contribution of the corresponding quarks.
Their specific content can be easily known by running the \texttt{help} command as shown in the following figure. 

\begin{figure}[h]
    \centering
    \includegraphics[width=1\linewidth]{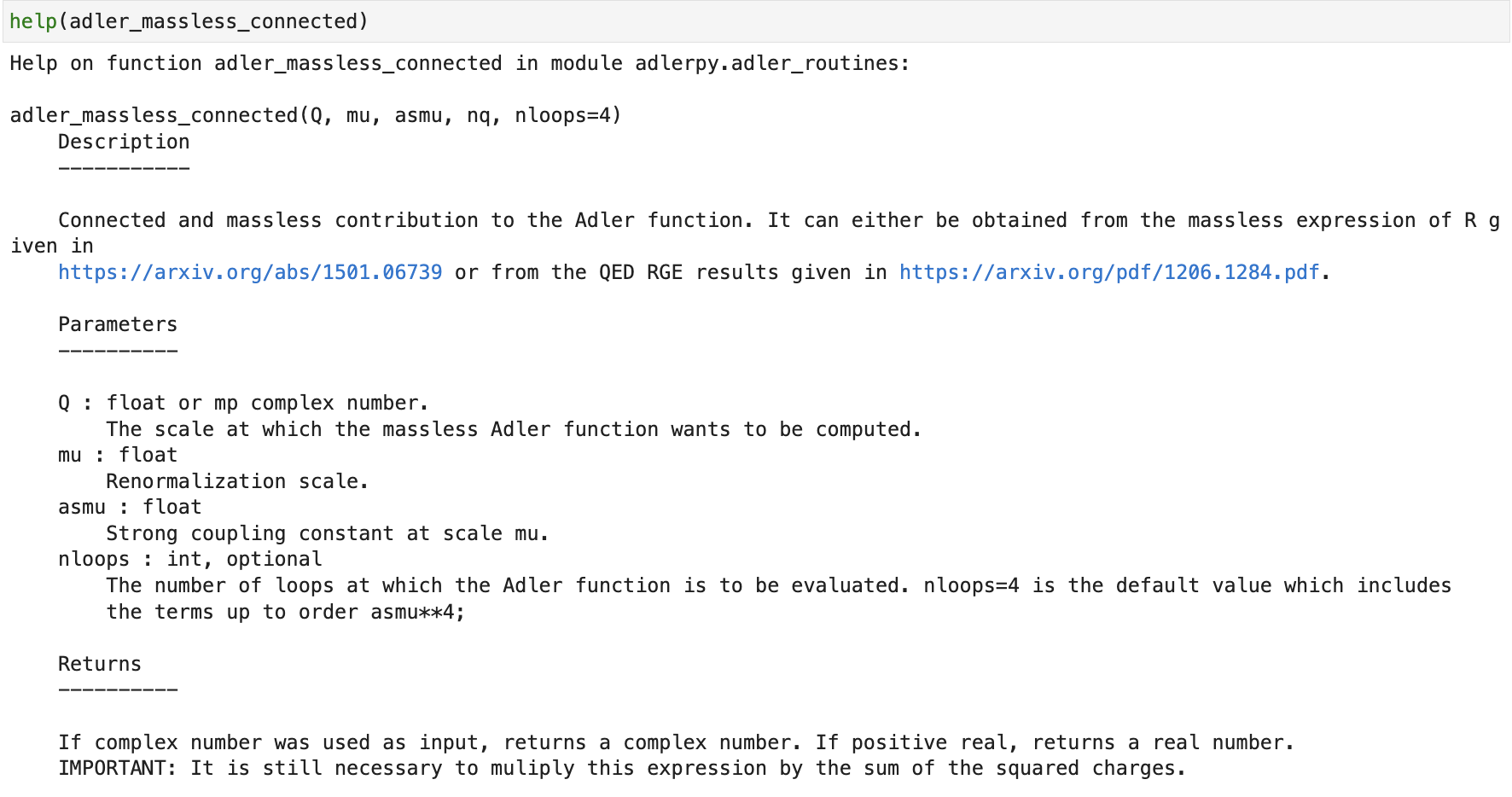}
    \caption{}
    \label{fig:codesnipethelp}
\end{figure}

\subsection*{Particle Class}

The code also features a \texttt{Particle} class used to represent particles. It includes attributes such as \texttt{name}, \texttt{mhat}, \texttt{mpole}, \texttt{mudec}, and \texttt{mpole\_on}, along with methods like \texttt{mrun}.

\bibliographystyle{unsrt}

\bibliography{example}

\end{document}